\documentclass[twocolumn,showpacs,prl,amsmath,amssymb]{revtex4}
\begin{document}

\title{Geometric derivation of quantum uncertainty}
\author{A. \surname{Kryukov}} 
\affiliation{Department of Mathematics, University of Wisconsin Colleges, 780 Regent Street, Madison, WI 53708} 
\date{\today}
\begin{abstract}
Quantum observables can be identified with vector fields on the sphere of normalized states. Consequently, the uncertainty relations for quantum observables become geometric statements. In the Letter the familiar uncertainty relation follows from the following stronger statement: Of all parallelograms with given sides the rectangle has the largest area.
\end{abstract}


\maketitle

%%%%%%%%%%%%%%%%%%%%%%%%%%%%%%%%%%%%%%%%%%%%%%%%%%%%%%%%%%%%%%%%%%%%%%%%%%%%%%%%%%%

Quantum observables can be identified with vector fields on the space of states. Namely, given a self-adjoint operator ${\widehat A}$ on a Hilbert space $L_{2}$ of square-integrable functions one can introduce the associated linear vector field $A_{\varphi}$ on $L_{2}$ by
\begin{equation}
\label{vector}
A_{\varphi}=-i{\widehat A}\varphi.
\end{equation}
This field is defined on a dense subset $D$ in $L_{2}$ on which the operator ${\widehat A}$ itself is defined.  Clearly, to know the vector field $A_{\varphi}$ is the same as to know the operator ${\widehat A}$ itself.
Moreover, the commutator of observables and the commutator (Lie bracket) of the corresponding vector fields are related in a simple way:
\begin{equation}
\label{comm}
[A_{\varphi},B_{\varphi}]=[{\widehat A},{\widehat B}]\varphi.
\end{equation}

The field $A_{\varphi}$ associated with an observable, being restricted to the sphere $S^{L_{2}}$ of unit normalized states, is tangent to the sphere. 
Indeed, the equation for the integral curves of $A_{\varphi}$ has the form
\begin{equation}
\label{SchroedA}
\frac{d \varphi_{\tau}}{d\tau}=-i{\widehat A}\varphi_{\tau}.
\end{equation}
The solution to (\ref{SchroedA}) through initial point $\varphi_{0}$ is given by 
$\varphi_{\tau}=e^{-i{\widehat A}\tau}\varphi_{0}$.
Here $e^{-i{\widehat A}\tau}$ denotes the one-parameter group of unitary transformations generated by $-i{\widehat A}$, as described by Stone's theorem.
It follows that the integral curve through $\varphi_{0} \in S^{L_{2}}$ will stay on the sphere. One concludes that, modulo the domain issues, the restriction of the vector field $A_{\varphi}$ to the sphere $S^{L_{2}}$ is a vector field on the sphere.

Under the embedding, the inner product on the Hilbert space $L_{2}$ gives rise to a Riemannian metric (i.e., point-dependent real-valued inner product) on the sphere $S^{L_{2}}$. For this one considers the realization $L_{2R}$ of the Hilbert space $L_{2}$, i.e., the real vector space of pairs $X=({\mathrm Re} \psi, {\mathrm Im} \psi)$ with $\psi$ in $L_{2}$. If $\xi, \eta$ are vector fields on $S^{L_{2}}$, 
%and $X=({\mathrm Re} \xi, {\mathrm Im} \xi)$, $Y=({\mathrm Re} \eta, {\mathrm Im} \eta)$, 
one can define a Riemannian metric $G_{\varphi}: T_{R\varphi}S^{L_{2}}\times T_{R\varphi}S^{L_{2}} \longrightarrow R$ on the sphere by
\begin{equation}
\label{Riem}
G_{\varphi}(X,Y)={\mathrm Re} (\xi, \eta).
\end{equation}
Here the tangent space $T_{R\varphi}S^{L_{2}}$ to $S^{L_{2}}$ at a point $\varphi$ is identified with an affine subspace in $L_{2R}$, $X=({\mathrm Re} \xi, {\mathrm Im} \xi)$, $Y=({\mathrm Re} \eta, {\mathrm Im} \eta)$ and $(\xi, \eta)$ denotes the $L_{2}$-inner product of $\xi, \eta$. 
%Note that whenever the inner product $(\xi,\eta)$ is real, the Riemannian and the $L_{2}$ inner products coincide.
Note that the obtained Riemannian metric $G_{\varphi}$ is {\em strong} in the sense that it yields an isomorphism ${\widehat G}:T_{R\varphi}S^{L_{2}}\longrightarrow \left (T_{R\varphi}S^{L_{2}}\right)^{\ast}$ of dual spaces.   

The Riemannian metric on $S^{L_{2}}$ yields a (strong) Riemannian metric on the projective space $CP^{L_{2}}$. For this one defines the metric on $CP^{L_{2}}$ so that the bundle projection $\pi: S^{L_{2}} \longrightarrow CP^{L_{2}}$ would be a Riemannian submersion. The resulting metric on $CP^{L_{2}}$ is called the Fubini-Study metric. To put it simply, 
an arbitrary tangent vector $X \in T_{R\varphi}S^{L_{2}}$
can be decomposed into two components: tangent and orthogonal to the fibre $\{\varphi\}$ through $\varphi$ (i.e., to the plane $C^{1}$ containing the circle $S^{1}=\{\varphi\}$). The differential $d\pi$ maps the tangent component to zero-vector. The orthogonal component of $X$ can be then identified with $d\pi(X)$. 
If two vectors $X,Y$ are orthogonal to the fibre $\{\varphi\}$, the inner product of $d\pi(X)$ and $d\pi(Y)$ in the Fubini-Study metric is equal to the inner product of $X$ and $Y$ in the metric $G_{\varphi}$. Note that the obtained Riemannian metrics on $S^{L_{2}}$ and $CP^{L_{2}}$ are invariant under the induced action of the group of unitary transformations on $L_{2}$.

Having a Riemannian metric on the manifolds $S^{L_{2}}$ and $CP^{L_{2}}$ opens a way for formulating the unitary and non-unitary processes in quantum mechanics in geometrical terms. Namely, as shown in Refs.\cite{Kryukov}-\cite{Kryukov2} (see also Ref.\cite{Kryukov3} for the mathematical considerations), both the Schr{\"o}dinger evolution and the process of collapse of a state can be thought of as geodesic motions on the sphere of states furnished with an appropriate strong Riemannian metric. Such a geometrization of quantum dynamics goes beyond the existing methods of geometrical quantum mechanics pioneered in Refs.\cite{Gun},\cite{Kib} (see Refs.\cite{Abbie}-\cite{Stulp} for extension of these ideas and review of other recent developments), and the geometric considerations related to Berry's phase (Refs.\cite{Berry}-\cite{Simon} amongst many others). Indeed, in those papers the metric on spaces of states is fixed and, consequently, is not dynamical. 

The goal of this work is to demonstrate that the more basic notions of expected value, variance and uncertainty relation also have a clear geometric interpretation. 
This interpretation is based directly on the association of observables with vector fields on the sphere of states and does not employ the Hamiltonian formalism on the phase space. This makes the interpretation particularly transparent and naturally leads one to a geometric uncertainty identity.

Let's begin with the standard uncertainty relation for observables ${\widehat A}, {\widehat B}$:
\begin{equation}
\label{uncert}
\Delta A \Delta B \ge \frac{1}{2}\left |\left(\varphi, [{\widehat A},{\widehat B}]\varphi\right)\right|.
\end{equation}
Here $\Delta A^{2}=(\varphi, {\widehat A}^{2}\varphi)-(\varphi, {\widehat A}\varphi)^{2}$ and similarly for $\Delta B^{2}$ and $\varphi$ is the state of the system under consideration. It is implicit in (\ref{uncert}) that the state $\varphi$ is in the domain of all operators involved.
As an immediate corollary of the relation one sees that, in general, 
the standard deviations $\Delta A, \Delta B$ of non-commuting observables cannot be made arbitrarily small at the same time (i.e., for the same state $\varphi$).
This constitutes a version of the famous uncertainty principle of Heisenberg \footnote {In the original Heisenberg formulation of the principle the issue of {\em simultaneous} measurements of observables is central. The mathematically derived uncertainty relations are not about such measurements, but rather about the standard deviations (or the like measures) of the observables measured separately on the same state.}. 

In light of the identification (\ref{vector}) of observables with vector fields on the sphere of states $S^{L_{2}}\subset L_{2}$, each term in (\ref{uncert}) obtains a simple geometric interpretation. Namely, the equality
\begin{equation}
{\overline A} \equiv (\varphi, {\widehat A}\varphi)=(-i\varphi, -i{\widehat A}\varphi),
\end{equation}
signifies that the expected value of an observable ${\widehat A}$ in the state $\varphi$ is the projection of the vector $-i{\widehat A}\varphi \in T_{\varphi}S^{L_{2}}$ on the vector $-i\varphi=-i I \varphi \in T_{\varphi}S^{L_{2}}$, associated with the identity operator $I$. 
Because
\begin{equation}
(\varphi, {\widehat A}^{2}\varphi)=({\widehat A}\varphi, {\widehat A}\varphi)=(-i{\widehat A}\varphi, -i{\widehat A}\varphi),
\end{equation}
the term $(\varphi, {\widehat A}^{2}\varphi)$ is just the norm of the vector $-i{\widehat A}\varphi$ squared. Note that the expected value $(\varphi, {\widehat A}_{\bot}\varphi)$ of the operator ${\widehat A}_{\bot} \equiv {\widehat A}-{\overline A}I$ in the state $\varphi$ is zero. Therefore, the vector $-i{\widehat A}_{\bot}\varphi=-i{\widehat A}\varphi-(-i{\overline A}\varphi)$, which is the component of $-i{\widehat A}\varphi$ orthogonal to $-i\varphi$  is  orthogonal to the entire fibre $\{\varphi\}$.
Accordingly, the variance 
\begin{equation}
\Delta A^{2}=(\varphi, ({\widehat A}-{\overline A}I)^{2}\varphi)=(\varphi, {\widehat A}_{\bot}^{2}\varphi)=(-i{\widehat A}_{\bot}\varphi, -i{\widehat A}_{\bot}\varphi) 
\end{equation}
is the norm squared of the component $-i{\widehat A}_{\bot}\varphi$. As discussed, the image of this vector under $d\pi$ can be identified with the vector itself. 
It follows that the norm of $-i{\widehat A}_{\bot}\varphi$ in the Fubini-Study metric coincides with its norm in the Riemannian metric on $S^{L_{2}}$ (and in the original $L_{2}$-metric). 

Consider the evolution equation
\begin{equation}
\label{evoll}
\frac{d\varphi_{t}}{dt}=-i{\widehat A}\varphi_{t}
\end{equation}
for the state $\varphi_{t}$ with the initial condition $\left.\varphi_{t}\right|_{t=0}=\varphi$.
By projecting both sides of this equation by $d\pi$, one obtains
\begin{equation}
\label{ddt}
\frac{d\{\varphi_{t}\}}{dt}=-i{\widehat A}_{\bot}\varphi_{t}.
\end{equation}
The left hand side of (\ref{ddt}) at $t=0$ is the velocity of evolution of the projection of $\varphi_{t}$ at the point $\{\varphi\}\in CP^{L_{2}}$. By the above, the norm of the right hand side at $t=0$ is the uncertainty of ${\widehat A}$ in the state $\varphi$:
\begin{equation}
\label{speed}
\|-i{\widehat A}_{\bot}\varphi\|=\Delta A.
\end{equation}
So the uncertainty $\Delta A$ is equal to the speed of the state $\{\varphi_{t}\}$ at the point $\{\varphi\}$ under the evolution (\ref{evoll}). In the case when ${\widehat A}$ is equal to the Hamiltonian ${\widehat h}$ of the system, one obtains the result of Ref.\cite{AA}: the energy uncertainty is the speed of evolution of the state in the projective space.
 
One concludes that the left hand side of (\ref{uncert}) is the product of norms of the projections of vectors $ -i{\widehat A}\varphi$, $-i{\widehat B}\varphi$ onto $T_{\{\varphi\}}CP^{L_{2}}$. In geometric terms, the left hand side is therefore the area $A_{|XY|}$ of a rectangle with sides of lengths $\|-i{\widehat A}_{\bot}\varphi\|$, $\|-i{\widehat B}_{\bot}\varphi\|$. Let's show that the right hand side of (\ref{uncert}) can be estimated via the area of parallelogram formed by vectors $-i{\widehat A}_{\bot}\varphi$, $-i{\widehat B}_{\bot}\varphi$.
For this note that
$[{\widehat A}, {\widehat B}]=[{\widehat A}_{\bot}, {\widehat B}_{\bot}]$
and, therefore,
\begin{eqnarray}
\label{im}
\nonumber
(\varphi, [{\widehat A}, {\widehat B}]\varphi)=
({\widehat A}_{\bot}\varphi, {\widehat B}_{\bot}\varphi) 
-({\widehat B}_{\bot}\varphi, {\widehat A}_{\bot}\varphi) \\
=2i {\mathrm Im}({\widehat A}_{\bot}\varphi, {\widehat B}_{\bot}\varphi) 
=2i {\mathrm Im}(-i{\widehat A}_{\bot}\varphi, -i{\widehat B}_{\bot}\varphi).
\end{eqnarray}
The form ${\mathrm Im}(\xi,\eta)$ is an anti-symmetric 2-form on vectors $\xi,\eta$. Let $\{e_{k}\}$ be an orthonormal basis in $L_{2}$, such that $e_{1}=-i{\widehat A}_{\bot}\varphi$ and the vector $-i{\widehat B}_{\bot}\varphi$ is in the linear envelop $C^{2}$ of the vectors
$e_{1}, e_{2}$. Let $E_{1}=e_{1}$, $E_{2}=ie_{1}$, $E_{3}=e_{2}$, $E_{4}=ie_{2}, \ ...$  be the corresponding orthonormal basis in the realization $L_{2R}$. Note that the linear envelop $R^{4}$ of the vectors $E_{1}, E_{2}, E_{3}, E_{4}$ is a subspace of the tangent space $T_{R\varphi}S^{L_{2}}$ and the Riemannian metric on the sphere yields the Euclidean metric on $R^{4}$. 
Let's denote the realization of the vectors $\xi=-i{\widehat A}_{\bot}\varphi, \eta=-i{\widehat B}_{\bot}\varphi$ by $X$ and $Y$ and let's denote the components of $X$ and $Y$ in the basis $\{E_{k}\}$ by $x_{k}$ and $y_{k}$ respectively. Because $x_{k}=y_{k}=0$ for $k>4$, one has
\begin{equation}
{\mathrm Im}(\xi,\eta)={\mathrm Im}\sum_{k}\xi_{k} {\overline \eta}_{k}=(x_{2}y_{1}-x_{1}y_{2})+(x_{4}y_{3}-x_{3}y_{4}),
\end{equation}
and so the right hand side of (\ref{uncert}) is equal to
\begin{equation}
\label{ima}
%\frac{1}{2}\left |\left(\varphi, [{\widehat A},{\widehat B}]\varphi\right)\right|=
 |(x_{1}y_{2}-x_{2}y_{1})+(x_{3}y_{4}-x_{4}y_{3})|.
\end{equation}
On the other hand, the area squared $A^{2}_{XY}$ of the parallelogram on vectors $X$, $Y$ is equal to
\begin{eqnarray}
\label{1n}
\nonumber
(x_{1}y_{2}-x_{2}y_{1})^{2}+(x_{1}y_{3}-x_{3}y_{1})^{2}+(x_{1}y_{4}-x_{4}y_{1})^{2}\\
+(x_{2}y_{3}-x_{3}y_{2})^{2}+(x_{2}y_{4}-x_{4}y_{2})^{2}+(x_{3}y_{4}-x_{4}y_{3})^{2}.
\end{eqnarray}
By the choice of $\{E_{k}\}$, we have $x_{2}=x_{3}=x_{4}=0$. By comparing (\ref{ima}) and (\ref{1n}) one concludes that
\begin{equation}
\label{areaEst}
A_{XY} \ge \frac{1}{2}\left |\left(\varphi, [{\widehat A},{\widehat B}]\varphi\right)\right|.
\end{equation}
As a result, the obvious geometric inequality 
\begin{equation}
\label{obv}
A_{|XY|} \ge A_{XY},
\end{equation}
implies the uncertainty relation (\ref{uncert}). 

It is well known that the uncertainty relation (\ref{uncert}) can be somewhat strengthened to take the form
\begin{equation}
\label{uncertS}
\Delta A^{2} \Delta B^{2} \ge \frac{1}{4}\left |\left(\varphi, [{\widehat A},{\widehat B}]\varphi\right)\right|^{2}+\frac{1}{4}\left |\left(\varphi, \{{\widehat A}_{\bot},{\widehat B}_{\bot}\}\varphi\right)\right|^{2},
\end{equation}
where $\{{\widehat A}_{\bot},{\widehat B}_{\bot}\}$ stands for the anticommutator of the operators ${\widehat A}_{\bot},{\widehat B}_{\bot}$. 
Note that 
\begin{eqnarray}
\label{re}
\nonumber
(\varphi, \{{\widehat A}_{\bot},{\widehat B}_{
\bot}\}\varphi)=
({\widehat A}_{\bot}\varphi, {\widehat B}_{\bot}\varphi)+({\widehat B}_{\bot}\varphi, {\widehat A}_{\bot}\varphi) \\
=2 {\mathrm Re}({\widehat A}_{\bot}\varphi, {\widehat B}_{\bot}\varphi)
=2 {\mathrm Re}(-i{\widehat A}_{\bot}\varphi, -i{\widehat B}_{\bot}\varphi).
\end{eqnarray}
So the second term on the right of (\ref{uncertS}) is simply the square of Riemannian inner product of vectors $-i{\widehat A}_{\bot}\varphi$, $-i{\widehat B}_{\bot}\varphi$. With the help of (\ref{im}) one can now identify the right hand side of (\ref{uncertS}) with
$|(-i{\widehat A}_{\bot}\varphi, -i{\widehat B}_{\bot}\varphi)|^{2}$.
Using (\ref{speed}), one concludes that (\ref{uncertS}) is simply the Cauchy-Schwarz inequality
\begin{equation}
\|-i{\widehat A}_{\bot}\varphi \|^{2}\|-i{\widehat B}_{\bot}\varphi\|^{2} \ge |(-i{\widehat A}_{\bot}\varphi, -i{\widehat B}_{\bot}\varphi)|^{2}
\end{equation}
for the vectors $-i{\widehat A}_{\bot}\varphi, -i{\widehat B}_{\bot}\varphi$.

Recall that the left hand side of the uncertainty relations (\ref{uncert}), (\ref{obv}), (\ref{uncertS}) is the product of lengths of vectors $X, Y$. In particular, in the basis $E_{k}$ one has:
\begin{equation}
\label{4}
\Delta A^{2} \Delta B^{2}=x^{2}_{1}(y^{2}_{1}+y^{2}_{2}+y^{2}_{3}+y^{2}_{4}).
\end{equation}
Note that the right hand sides of the uncertainty relations (\ref{uncert}), (\ref{obv}) and (\ref{uncertS}) are formed by the terms of (\ref{4}). In particular, these uncertainty relations follow from (\ref{4}).
Moreover, the right hand side of (\ref{4}) is exactly the sum of the Riemannian inner product term squared
$\left ({\mathrm Re} (-i{\widehat A}_{\bot}\varphi, -i{\widehat B}_{\bot}\varphi)\right)^{2}=G^{2}_{\varphi} (X, Y)=x^{2}_{1}y^{2}_{1}$
and the area term squared
$A^{2}_{XY}=x^{2}_{1}(y^{2}_{2}+y^{2}_{3}+y^{2}_{4})$.
It follows that the uncertainty relation can be written in the form of the ``uncertainty identity''
\begin{equation}
\label{Pyth}
\Delta A^{2} \Delta B^{2}=A^{2}_{XY}+ G^{2}_{\varphi} (X, Y),
\end{equation}
with $X=-i{\widehat A}_{\bot}\varphi$ and $Y=-i{\widehat B}_{\bot}\varphi$.

One concludes, once again, that $A_{XY}=0$ is a necessary condition for vanishing uncertainty $\Delta A \Delta B$. 
This condition is satisfied when vectors $-i{\widehat A}_{\bot}\varphi$ and $-i{\widehat B}_{\bot}\varphi$ are linearly dependent over ${\mathrm R}$. Another necessary condition that follows from  (\ref{Pyth}) is the condition of orthogonality of the vectors $-i{\widehat A}_{\bot}\varphi$ and $-i{\widehat B}_{\bot}\varphi$ in the Riemannian metric. The necessary and sufficient condition for $\Delta A \Delta B=0$ is the vanishing of both terms on the right hand side of (\ref{Pyth}).
In particular, for bounded operators ${\widehat A}, {\widehat B}$, the uncertainty $\Delta A \Delta B$ vanishes iff at least one of the vectors $-i{\widehat A}_{\bot}\varphi$, $-i{\widehat B}_{\bot}\varphi$ vanishes. That is, iff $\varphi$ is an eigenstate of either ${\widehat A}$ or ${\widehat B}$. For example, for the Pauli matrices, $\Delta \sigma_{x} \Delta \sigma_{y}=0$ iff $\varphi$ is an eigenstate of either ${\widehat \sigma}_{x}$ or ${\widehat \sigma}_{y}$. 

Assume now that $[{\widehat A}, {\widehat B}]=cI$, where $c$ is a number. Recall that $A_{XY} \ge \frac{1}{2}\left|\left(\varphi, [{\widehat A}, {\widehat B}] \varphi \right) \right|$ and so the first term on the right of (\ref{Pyth}) is at least $|c/2|$. Therefore, the uncertainty $\Delta A \Delta B$ is at least $|c/2|$. This minimal value of the uncertainty can only be achieved if $A_{XY}=|c/2|$ and  $G_{\varphi} (X, Y)=0$. Recall that in the basis $E_{k}$ one has $A^{2}_{XY}=x^{2}_{1}\left(y^{2}_{2}+y^{2}_{3}+y^{2}_{4}\right)$ and $\frac{1}{2}\left|\left(\varphi, [{\widehat A}, {\widehat B}] \varphi \right) \right|=|x_{1}y_{2}|$. Therefore, to achieve the minimum value one must have $y^{2}_{3}+y^{2}_{4}=0$. It follows that $-i{\widehat B}\varphi=\lambda \left (-i{\widehat A}\varphi \right )$ for some complex $\lambda$. The condition $G_{\varphi} (X, Y)=0$ reads in the basis $E_{k}$ as $x_{1}y_{1}=0$. It follows that $y_{1}$ must be zero, which means that the constant $\lambda$ is purely imaginary.
In particular, for the momentum and position operators ${\widehat p}$ and ${\widehat x}$ these conditions yield Gaussian states for which $\Delta p \Delta x =\hbar /2$. 

Note that the terms on the right of (\ref{Pyth}) can be written as  
$\left \|X\right\|^{2}\left \|Y\right\|^{2}\sin^{2}\theta$ and $\left \|X\right\|^{2}\left \|Y\right\|^{2}\cos^{2}\theta$, where $\theta$ is the angle between the vectors $X$ and $Y$. In particular, when  $\theta=0$ the uncertainty comes from the inner product term $G_{\varphi} (X, Y)$ only and when $\theta=\pi/2$, the uncertainty is due to the area term. By replacing ${\widehat B}$ with a real linear combination of the operators ${\widehat A}$, ${\widehat B}$, one can change $\theta$ in any desirable way while preserving the uncertainty $\Delta A \Delta B$.

The standard uncertainty relations (\ref{uncert}), (\ref{uncertS}), the derived geometric uncertainty relation (\ref{obv}) and the uncertainty identity (\ref{Pyth}) are mathematical statements.
The mystery of the uncertainty principle lies not so much in these statements, but rather in a physical interpretation of operators and states entering the statements. 
So, what is the significance of the provided derivation in this respect?

The quantum evolution of a system yields a path on the sphere of states. The projection $\pi: S^{L_{2}} \longrightarrow CP^{L_{2}}$ gives then a path on the projective space $CP^{L_{2}}$ of physical states.
As advocated in Refs.\cite{Kryukov}-\cite{Kryukov2}, the evolution of state along the manifolds $S^{L_{2}}$ and $CP^{L_{2}}$  should be treated as a fundamental physical process, rather than just a way of describing changes in probability distributions of measured quantities. 
As shown in Ref.\cite{Kryukov2}, by choosing an appropriate Riemannian metric on the sphere $S^{L_{2}}$, one can ensure that the Schr{\"o}dinger path of the state is a geodesic on the sphere. 
Moreover, at least in the finite dimensional spaces of states, the process of collapse can be also modeled by a geodesic motion of the state in the metric perturbed by the measuring device.
The Born rule for probability of collapse can be derived from simple additional assumptions (see Ref.\cite{Kryukov2}). 

One is faced then with a new point of view on quantum mechanics that makes that theory quite similar to Einstein's general relativity, but considered on a manifold of states rather than on space-time. The approach turns out to be fruitful in explaining various paradoxical results in quantum theory via the geometry of the manifold of states. Moreover, the formalism allows one to naturally embed the physics of macroscopic particles on the classical Riemannian space into the theory (see Ref.\cite{Kryukov}).
In light of this, the provided geometric derivation of the uncertainty relation and the uncertainty identity seems to be another piece of the puzzle falling into place. 

What is the physical interpretation of quantum uncertainty in the the new geometrical setting?
The answer depends on the one's definition of the uncertainty. Here are some possible definitions together with their geometric interpretation. 

($\alpha$)\  Note first of all that the set of eigenstates of two non-commuting observables ${\widehat A}, {\widehat B}$ form two non-identical (often, non-overlapping) subsets $S_{A}, S_{B}$ of the sphere of states. 
If the intersection $S_{A}\cap S_{B}$ is empty, the state cannot belong to both of them at once. If the state is close in the Riemannian metric to one of these subsets, it cannot be arbitrarily close to the other one, hence, the uncertainty principle. 

Mathematically, the principle can be formulated in this case via the triangle inequality on the sphere of states. Namely, if $\varphi$  is the state of the system and $d(\varphi, S_{A})$, $d(\varphi, S_{B})$, $d(S_{A}, S_{B})$ are the distances in the Riemannian metric between $\varphi$ and  $S_{A}$, $\varphi$ and $S_{B}$, $S_{A}$ and $S_{B}$ respectively, then 
\begin{equation}
\label{uncert-dAB}
d(\varphi, S_{A})+d(\varphi, S_{B})\ge d(S_{A}, S_{B}).
\end{equation}
By projecting on $CP^{L_{2}}$, one obtains a similar inequality for physical states.  

In such an interpretation the uncertainty of an observable ${\widehat A}$ is the {\em distance} from the state to the set of eigenstates of ${\widehat A}$ in the Riemannian metric. 
The uncertainty relation (\ref{uncert-dAB}) shows that for two observables with no common eigenvectors the state cannot be made arbitrarily close to both $S_{A}$ and $S_{B}$ at once. For example, for spin states $\varphi$ of a non-relativistic electron one has $d(\{\varphi\}, \{S_{ \sigma_{x}}\})+d(\{\varphi\}, \{S_{ \sigma_{y}}\})\ge \frac{\pi}{2}$.

($\beta$)\  More commonly, the uncertainty of an observable ${\widehat A}$ in state $\varphi$ is defined as the {\em standard deviation} $\Delta A$. Recall that $\Delta A$ is the norm of the velocity vector $-i{\widehat A}_{\bot}\varphi$
of the evolution $\frac{d\{\varphi_{t}\}}{dt}=-i{\widehat A}_{\bot}\varphi_{t}$.
The velocity vector vanishes at the eigenstates (and only at the eigenstates) of the operator ${\widehat A}$. Therefore, the uncertainty $\Delta A$ vanishes only at the eigenstates as well. 

Note that in the case of the space $CP^{1}$ of spin states of a non-relativistic electron, the standard deviation $\Delta A$ of any observable ${\widehat A}$ with  $-i {\widehat A} \in su(2)$ can be identified with the distance $d(\{\varphi\}, \{S_{A}\})$  between the state and the set of eigenstates of ${\widehat A}$ (see Ref.\cite{Kryukov2}). In other words, the speed of evolution of the state in $CP^{1}$ is proportional to the distance $d(\{\varphi\}, \{S_{A}\})$.  In this particular case the definitions ($\alpha$) and ($\beta$) coincide. 

($\gamma$) \ The uncertainty can be understood as the product $\Delta A \Delta B$ of standard deviations of two observables for a system in a given state $\varphi$ (or, in some cases, as the infimum of the set of such products for all possible states). Suppose that the velocity vectors $-i{\widehat A}_{\bot}\varphi$, $-i{\widehat B}_{\bot}\varphi$, considered as vectors in the real space $L_{2R}$, are linearly dependent. Then the area of the parallelogram based on these vectors vanishes. In this case the right hand side of the geometric uncertainty relation (\ref{obv}) also vanishes. This provides one with a simple geometrical {\em necessary} condition for vanishing $\Delta A \Delta B$. 

($\delta$) \ A related and most common understanding of quantum uncertainty is based on the standard uncertainty relation (\ref{uncert}). 
This relation is often used to identify quantum uncertainty in the sense ($\gamma$) with non-commutativity of quantum observables under consideration. 
Note however that according to (\ref{uncertS}),
the lower bound of the product of standard deviations of two commuting observables on a given set of states may be positive \footnote{Of course, the lower bound of the product of standard deviations of two commuting observables over the {\em entire} sphere of states is zero. However, it is an unnecessary limitation to consider this case only.}. Conversely, even if two observables do not commute, they could still have a common eigenvector so that the standard deviations of both observables on this vector would vanish. In other words, the non-commutativity of observables ${\widehat A}, {\widehat B}$ is neither necessary nor sufficient for a nontrivial uncertainty relation. 

It is a pleasure to thank Malcolm Forster for numerous discussions that helped shaping this paper.

\end{document}